\documentclass[10pt, conference]{ieeeconf}      
\IEEEoverridecommandlockouts                              
\usepackage{cite}\usepackage{hyperref}
\usepackage{amsmath,amssymb,amsfonts}
\usepackage{graphicx}
\usepackage{textcomp}
\usepackage{amsmath,amssymb,bm,bbm,mathrsfs,amscd}
\usepackage{calc}
\usepackage{color}
\usepackage{xcolor}

\usepackage{dsfont}
\usepackage{graphicx}
\usepackage{epstopdf}
\usepackage{epsfig}
\usepackage{tikz}
\usepackage{amsmath}
\usepackage{amsfonts}
\usepackage{amssymb}
\usepackage{rotating}
\usepackage{mathtools}
\usepackage{color}
\usepackage{subfig}
\usepackage{enumerate,pgfplots}
\usepackage{algorithm}
\usepackage{algpseudocode}
\newtheorem{theorem}{Theorem}

\newtheorem{proposition}{Proposition}

\newtheorem{assumption}{Assumption}
\newcommand{\ba}{\begin{array}}
\newcommand{\ea}{\end{array}}

\newcommand{\be}{\begin{equation}}
\newcommand{\ee}{\end{equation}}

\newcommand{\mc}{\mathcal}

\newcommand{\ov}{\overline}

\newcommand{\Z}{\mathbb{Z}}

\def\1{\mathds{1}}

\newcommand{\R}{\mathbb{R}}

\newcommand{\V}{\mathcal{V}}

\newcommand{\G}{\mathcal{G}}

\def\MA#1{\textcolor{black}{#1}}

\DeclareMathOperator*{\argmin}{argmin}

\def\Z{\mathbb{Z}}

\def\R{\mathbb{R}}

\def\0{\boldsymbol{0}}

\tikzstyle{v_c}=[circle, draw,inner sep=2pt, minimum width=12pt, color=blue]
\tikzstyle{v_a}=[circle, draw,inner sep=2pt, minimum width=12pt, color=red]
\tikzstyle{edge} = [draw,thick,-,font=\small ]
\tikzstyle{label} = [draw,fill=black,font=\normalsize]

\def\G{{\mathcal G}}

\def\BibTeX{{\rm B\kern-.05em{\sc i\kern-.025em b}\kern-.08em
	T\kern-.1667em\lower.7ex\hbox{E}\kern-.125emX}}
	\title{\LARGE \bf Model Predictive Control for Coupled Adoption-Opinion Dynamics}
	\author{Martina~Alutto, Qiulin Xu, Fabrizio Dabbene, Hideaki Ishii and Chiara Ravazzi 
\thanks{Martina~Alutto, Fabrizio Dabbene and Chiara Ravazzi are with the Institute of Electronics, Computer and Telecommunication Engineering, National Research Council of Italy, Politecnico di Torino, 10129 Torino, Italy (e-mail: {\{martinaalutto;\,chiararavazzi;\,fabriziodabbene\}@cnr.it}).
	Qiulin Xu is with the Department of Computer Science, Institute of Science Tokyo, Yokohama 226-8502, Japan (e-mail: xu.q.76b5@m.isct.ac.jp).
	Hideaki Ishii is with the Department of Information Physics and Computing, The University of Tokyo, Tokyo 113-8654, Japan (e-mail:hideaki\_ishii@ipc.i.u-tokyo.ac.jp).
}
\thanks{This work has been supported by the European Union -- Next Generation EU, Mission 4, Component 1, under the PRIN project {\em{TECHIE: A control and network-based approach for fostering the adoption of new technologies in the ecological transition}}, Cod. 2022KPHA24, CUP Master: D53D23001320006, CUP: B53D23002760006, and by Japan Science and Technology Agency as part of the ASPIRE program under Grant No. JPMJAP2402.}%
}

\begin{document}
\thinmuskip=0.3mu  
\medmuskip=0.3mu   
\thickmuskip=0.3mu 

\maketitle
\thispagestyle{empty}
\pagestyle{empty}

\begin{abstract}	
	This paper investigates an optimal control problem for an adoption-opinion model that couples opinion dynamics with a compartmental adoption framework on a multilayer network to study the diffusion of sustainable behaviors. Adoption evolves through social contagion and perceived benefits, while opinions are shaped by social interactions and feedback from adoption levels. Individuals may also stop adopting virtuous behavior due to external constraints or shifting perceptions, affecting overall diffusion. After the stability analysis of equilibria, both in the presence and absence of adopters, we introduce a Model Predictive Control (MPC) framework that optimizes interventions by shaping opinions rather than directly enforcing adoption. This nudge-based control strategy allows policymakers to influence diffusion indirectly, making interventions more effective and scalable. Numerical simulations demonstrate that, in the absence of control, adoption stagnates, whereas MPC-driven interventions sustain and enhance adoption across communities.
\end{abstract}

\section{Introduction} 
The transition to sustainable lifestyles is one of the defining challenges of the 21st century, requiring individuals, communities, and policymakers to adopt behaviors that reduce environmental impact, such as lowering energy consumption and embracing sustainable mobility \cite{Steg2018}. 
Despite growing awareness of climate change and the availability of green technologies, psychological, economic, and social barriers often hinder the diffusion of sustainable practices \cite{Kollmuss2002}.
However, behavioral science and network theory suggest that social influence plays a crucial role in shaping individual decisions \cite{Cialdini2003}, making the study of social contagion and opinion dynamics essential for understanding and accelerating green transitions.

In recent years, mathematical models have been extensively used to analyze the innovation diffusion, with applications in marketing strategies, adoption of electric vehicles and renewable energy technologies. Classical models, such as the Bass diffusion model \cite{Bass1969}, describe adoption as a mix of individual decision-making and peer influence. However, adoption is inherently a social process, strongly shaped by interactions within a network. To better capture these dynamics, network-based models \cite{BRESCHI2023103651,VILLA2024106106} describe adoption as a process driven by social reinforcement and individual attitude. The spread of sustainable behaviors can thus be understood within the broader class of diffusion processes, which also encompasses phenomena such as epidemic outbreaks in interconnected populations or the uptake of new technologies. Despite their different domains, these dynamics share a common feature: they are driven by simple local mechanisms on complex networks, namely pairwise interactions among individuals and spontaneous changes in state. 
	
Within this broad spectrum of approaches, agent-based models allow for a fine-grained representation of individual behaviors, whereas compartmental models, commonly used in epidemiology \cite{Hethcote2000}, are more suitable for representing adoption \cite{Kumar2022, dhar2015impact} at a macroscopic level, by grouping individuals into compartments that capture the different stages of diffusion.
In particular, the Susceptible-Infected-Vigilant (SIV) \cite{Xu2024} or Susceptible-Infected-Recovered-Susceptible (SIRS) \cite{LI20141042,ZHANG2021126524} model incorporate temporary immunity, from both passive recovery from infection and active preventive measures before infection. This can be interpreted in the context of green innovation adoption as individuals who temporarily adopt a behavior but may temporarily abandon it due to external factors such as economic constraints or shifting social influences. 
One crucial difference between epidemic control and green behavior adoption lies in the policy objectives. In epidemiology, control strategies are designed to minimize the infection spread \cite{acemoglu2021optimal}.
In contrast, sustainability policies seek to maximize the spread of pro-environmental behaviors, ensuring that sustainable practices reach a critical mass within the population. This fundamental distinction highlights the need for tailored intervention strategies that leverage social influence, incentives, and network effects to accelerate the adoption of green innovations.

While epidemic models provide a useful framework for behavioral spread, they often assume individuals are uniformly susceptible to influence. In reality, decision-making is shaped not only by exposure to new behaviors but also by opinion formation and social influence. 
Building on this rationale, we first propose a novel two-layer adoption-opinion model that combines a compartmental adoption framework (Susceptible–Adopter–Dissatisfied, SAD) with opinion dynamics, closely resembling the epidemic model studied in \cite{Xu2024}. This formulation captures not only exposure-driven adoption but also heterogeneity in individual predispositions and the persistence of attitudes over time. Unlike \cite{Xu2024}, we draw on the Friedkin–Johnsen model of opinion dynamics \cite{Friedkin1990}, which balances intrinsic predisposition with social influence. This allows us to represent stubborn individuals and networks that are not polarized, that is an essential feature when modeling the adoption of green behaviors, where variability in opinions and willingness to adopt is significant. We consider adoption and opinion processes on distinct but interconnected layers, a common approach in the literature \cite{Paarporn2017, She2022, Wang2022}. The physical layer governs adoption through pairwise encounters, where the probability of adopting a service or technology increases with the observed diffusion within one’s community and neighboring communities, reflecting the collective perception of quality, reliability, and maturity. The social layer captures opinion evolution through inter-community interactions and feedback from observed adoption. Although the two layers may appear largely aligned, their separation is crucial to distinguish mechanisms of influence operating through different channels.
We then study the equilibria of the dynamics, identifying both the adoption-free and the adoption-diffused equilibrium points, and we perform a stability analysis to characterize the conditions under which each equilibrium arises.

Finally, we address an optimal control problem aimed at maximizing the number of adopters of sustainable behaviors. Rather than enforcing adoption directly, we study policies that shape opinions, which in turn modulate adoption dynamics. This two-layer control strategy enables policymakers to influence diffusion indirectly, making interventions more effective and scalable. To achieve this, we develop a nonlinear Model Predictive Control (MPC) framework \cite{Rawlings2014}, which enables us to design optimal intervention strategies while accounting for realistic constraints, such as limited policy budgets and the dynamic nature of the adoption process. We further compare the performance of the MPC approach with a constant, non-adaptive policy, highlighting the advantages of feedback-based, adaptive control in achieving sustained long-term behavior change.

The rest of the paper is organized as follows. In Section \ref{sec:model} we introduce the adoption-opinion model. Stability results are provided in Section \ref{sec:stability}, while the optimal control problem we address is presented in Section \ref{sec:mpc}. In Section \ref{sec:conclusion}, we discuss future research lines.
\MA{All proofs can be found in \cite{alutto2025preprint}.}

\subsection{Notation}
We denote by $\R$ and $\R_{+}$ the sets of real and nonnegative real numbers. 
The all-1 vector and the all-0 vector are denoted by $\1$ and $\0$ respectively. The identity matrix and the all-0 matrix are denoted by $I$ and $\mathbb{O}$, respectively.
The transpose of a matrix $A$ is denoted by $A^T$. 
For $x$ in $\R^n$, let $||x||_1=\sum_i|x_i|$ and $||x||_{\infty}=\max_i|x_i|$ be its $l_1$- and $l_{\infty}$- norms, while $\mathrm{diag}(x)$ denotes the diagonal matrix whose diagonal coincides with $x$. 
For an irreducible matrix $A$ in $\R_+^{n\times n}$, we let $\rho(A)$ denote the spectral radius of $A$. 
Inequalities between two matrices $A$ and $B$ in $\mathbb{R}^{n \times m}$ are understood to hold entry-wise, i.e., $A \leq B$ means $A_{ij} \leq B_{ij}$ for all $i$ and $j$, while $A < B$ means $A_{ij} < B_{ij}$ for all $i$ and $j$. 

\section{Model Description}\label{sec:model}
We consider a closed population, with no inflow or outflow of people, divided into different communities $\mathcal{V}=\{1,\ldots,n\}$, representing groups characterized by socio-demographic aspects such as age, educational level, geographic area, or other relevant characteristics. Within each community $j\in\mathcal{V}$, individuals are further classified into three dynamic compartments, each represented by a fraction of the community: 
\begin{itemize}
	\item $s_j(t)$ represents the fraction of susceptible individuals, that is, people who are not adopting a certain virtuous behavior or service/technology at time $t\in\Z_+$, but may do so in the future;
	\item $a_j(t)$ is the fraction of adopters at time $t\in\Z_+$;
	\item $d_j(t)$ represents the fraction of dissatisfied individuals at time $t\in\Z_+$, i.e., those who have adopted in the past but are not satisfied and temporarily abandon it, or those who are not convinced that adoption is beneficial and therefore do not intend to adopt it in the near future.
\end{itemize}

\begin{figure}
	\hspace{-0.3cm}
	\subfloat[]{\includegraphics[scale=0.75]{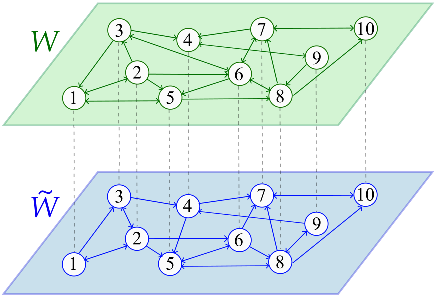}}
	\subfloat[]{\includegraphics[scale=0.4]{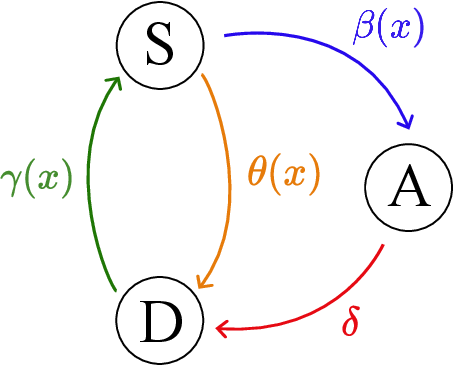}}
	\caption{(a) The bilayer network of the coupled adoption-opinion model. (b) Adoption model with three states and various transition parameters.}
	\label{fig:layers}
\end{figure}

We assume that each community has a certain opinion $x_j(t)\in[0,1]$ about the virtuous behavior at time $t\in\Z_+$.
Our model is based on a two-layer network, as shown in Figure \ref{fig:layers}(a). The lower layer represents the social network for opinion evolution, influenced by social imitation and prevailing opinions; while the upper layer models the physical interaction network. 

\subsection{Adoption dynamics}

We model the physical network as a finite weighted directed graph $\mc G=(\mc V, \mc E, W)$, where $\mc V = \{1, 2, \dots, n\}$ is the set of nodes associated to each community, $\mc E \subseteq \mc V \times \mc V$ is the set of directed links, and $W \in \mathbb{R}_+^{n \times n}$ is a nonnegative weight matrix, known as the physical interaction matrix, such that $W_{jk} > 0$ if and only if there is a link $(j,k) \in \mc E$ from node $j$ to node $k$. The physical neighborhood of a node $j \in \mc V$ is defined as $\mathcal{N}_j = \{k \in \mc V : (j, k) \in \mc E\}$.
The adoption process among $n$ interacting communities is described by the following discrete-time equations:
\begin{align}\label{eq:adoption-model}
	s_j(t+1) &=s_j(t) -\beta_j(x_j(t)) s_j(t)\displaystyle\sum_{k \in \mc N_j} W_{jk} a_k(t)+ \nonumber \\
	& \qquad+\gamma_j(x_j(t)) d_j(t) - \theta_j(x_j(t)) s_j(t)\,, \\[1pt]
	a_j(t+1) &= a_j(t) + \beta_j(x_j(t)) s_j(t)\!\! \displaystyle\sum_{k \in \mc N_j} W_{jk} a_k(t) - \delta_j a_j(t) \,,\nonumber \\[1pt]
	d_j(t+1) &= d_j(t) - \gamma_j(x_j(t))d_j(t) \!+\! \theta_j(x_j(t))s_j(t) + \delta_j a_j(t) \,,\nonumber
\end{align}
for all $j \in \V$, where $\beta_j : [0,1] \rightarrow [0,1]$ is the adoption susceptibility rate, $\gamma_j: [0,1] \rightarrow [0,1]$ is the reversion rate (probability of dissatisfied individuals becoming susceptible), $\theta_j: [0,1] \rightarrow [0,1]$ is the unsatisfaction rate, and $\delta_j \in [0,1]$ is the dismissing rate for community $j\in\mathcal{V}$.
A community holding a favorable opinion is more likely to adopt green behaviors and less likely to remain or become dissatisfied.
For simplicity, in this work we will assume $\beta_j(x_j) = \beta_j x_j$, $\gamma_j(x_j) = \gamma_j x_j$, $\theta_j(x_j) = \theta_j (1 - x_j)$ but the model can be extended to any choice of increasing functions for $\beta_j$ and $\gamma_j$, and decreasing for $\theta_j$ in $x_j$. 
\MA{These assumptions make the model analytically tractable while still capturing the key interplay between opinion and adoption dynamics.}
Figure \ref{fig:layers}(b) shows the three-state adoption model where the transitions between different compartments are shown with arrows. 


\subsection{Opinion dynamics}
Each community $j$ is endowed with an opinion variable $x_j \in [0,1]$, which represents the average belief of the community regarding the innovation/service. Each community continuously reassesses its opinion by comparing it with the opinions of its neighbors. We model the interactions through social networks (e.g. local community networks, familiar networks, or virtual social networks, forums discussion networks) using a directed graph $\tilde{\mc G}= (\mc V, \tilde{\mc E}, \tilde{W})$, where $\tilde{\mc E} \subseteq \mc V \times \mc V$ is the set of directed links and $\tilde{W} \in \R_+^{n \times n}$ is a nonnegative weight matrix, referred to as the social interaction matrix. \MA{The social neighborhood of a node $j \in \mc V$ is defined as $\tilde{\mathcal{N}}_j = \{k \in \mc V : (j, k) \in \tilde{\mc E} \}$.}
In this context, we extend the Friedkin-Johnsen model of opinion dynamics \cite{Friedkin1990}, and assume
\be\label{eq:opinion-model} x_j(t+1) =\alpha_j x_j(0) + \lambda_j  \mspace{-6mu} \sum_{k \in \tilde{\mc N}_j}\tilde{W}_{jk}x_k(t) + \xi_j  \mspace{-6mu} \displaystyle\sum_{k \in \mc N_j}W_{jk}a_k(t)\,,
\ee
\MA{where $x_j(0)$ represents the initial opinion of community $j\in\V$ and $\alpha_j +\lambda_j+\xi_j= 1$ balance the intensity of intrinsic predisposition, social interactions, and observed adoption. Specifically, 
quantifies the susceptibility of the community to its neighbors’ opinions, while $\xi_j$ captures the weight of adoption-driven feedback from the physical layer. The novelty of this formulation lies in combining social influence and adoption feedback within the opinion update. Unlike standard models, our approach updates opinions based on both neighbors’ beliefs and observed adoption, capturing realistic diffusion with persistent heterogeneity. Purely social models may overestimate convergence or miss these patterns.}
\subsection{Coupled adoption-opinion model}
Notice that the models \eqref{eq:adoption-model} and \eqref{eq:opinion-model} are coupled.
The next assumption guarantees that the two underlying networks are connected.
\begin{assumption}\label{ass:ass1}
	Assume that both $W$ and $\tilde{W}$ are row-stochastic and irreducible. Moreover, assume that 
	$\gamma_j + \theta_j \in (0,1)$
	for all $j \in\V$.
\end{assumption}\smallskip

The next result ensures the well-posedness of the adoption-opinion model. 
\begin{proposition}\label{prop:invariant}
	Consider the adoption-opinion model \eqref{eq:adoption-model}-\eqref{eq:opinion-model} under Assumption \ref{ass:ass1}. Then, if $s(0), a(0), d(0)$ in $[0,1]^{\V}$ and $s(0)+ a(0)+ d(0) = \1$, then $s(t), a(t),d(t)$ in $[0,1]^{\V}$ and $s(t)+a(t)+d(t)=\1$ for all $t\geq0$. Moreover, if $x(0)$ in $[0,1]^{\V}$, then $x(t)$ in $[0,1]^{\V}$ for all $t \geq 0$.
\end{proposition}\smallskip

As a consequence of Proposition \ref{prop:invariant}, just two state variables and opinions are sufficient to describe the configuration of the adoption. From now on, we will consider the following dynamics in vectorial form,
\begin{align}\label{eq:vector_model} 
	& a(t\! +\!1) \!\!=(I \!\!+\!\! B \mathrm{diag}(x(t)) \!\!\mathrm{diag}(\1\! -\! a(t)\!-\! d(t)) W\! -\! \Delta )a(t),\nonumber \\
	&d(t+1) \!\!= d(t) - \Gamma \mathrm{diag}(x(t)) d(t)  + \Delta a(t)+\nonumber \\
	&\qquad\quad+ \Theta (I - \mathrm{diag}(x(t))) (\1 -a(t)- d(t))\,, \\
	&x(t+1) \!\!= (I-\Lambda-\Xi) x(0) + \Lambda \tilde{W} x(t) + \Xi W a(t),\nonumber
\end{align}
where $\Delta =  \mathrm{diag}(\delta)$, $B =  \mathrm{diag}(\beta)$, $\Gamma =  \mathrm{diag}(\gamma)$, $\Theta =  \mathrm{diag}(\theta)$, $\Lambda=\mathrm{diag}(\lambda)$, $\Xi=\mathrm{diag}(\xi)$.\medskip

\section{Stability results}\label{sec:stability}
In this section we study the equilibria of the system in \eqref{eq:vector_model}. Furthermore, we analyze stability conditions and discuss the mutual influence between the spread of innovation and the evolution of opinions. 
To this end, we pose the following assumption related to the social interaction network, in order to have that each community is influenced, even indirectly, by at least one communities with some stubbornness. 
\begin{assumption}\label{ass:ass2}
	For any node $j\in\V$ there exists a path in $\widetilde{\G}$ from $j$ to $k$ with $\alpha_k>0$. 
\end{assumption}

\MA{In other words, every community is either stubborn itself or indirectly connected to at least one stubborn community.}

\subsection{Adoption-free equilibrium}
Define $\Psi(x) := \big( (\Gamma- \Theta)\mathrm{diag}(x) + \Theta\big)^{-1}\Theta(I-\mathrm{diag}\big(x\big)).$
The following proposition guarantees that there exists an adoption-free equilibrium, i.e. a stationary configuration with no adopters.
\begin{proposition}
	Given Assumption \ref{ass:ass2}, 
	$\big(\0,\Psi(x^*)\1, x^* \big)$ is an equilibrium for \eqref{eq:vector_model}
	with
	$x^*=(I- \Lambda\tilde{W})^{-1} (I-\Lambda-\Xi) x(0)$.
\end{proposition}\smallskip

A key concept in epidemic models is the reproduction number, which indicates the expected number of secondary cases generated by a single infected individual in a susceptible population. Analogously, in the context of innovation diffusion, this metric can be seen as the threshold determining whether a new green behavior will achieve widespread adoption or remain confined to a small group of early adopters. 
For the system in \eqref{eq:vector_model}, we define the opinion-dependent reproduction number as
\begin{equation*}
	R_0^A(x(t)) = \rho\left(I - \Delta +B \mathrm{diag}(x(t))\left(I - \Psi(x(t))\!\!\right)\!W\!\right)\,.
\end{equation*}
This varies as a function of the opinion states $x(t)$, reflecting the dynamic influence of social beliefs on the adoption process. The next results provide conditions for the stability of the adoption-free equilibrium. 
Consider first a scenario when all communities strongly support the innovation, which yields
\begin{equation}\label{eq:r0-max}
	R_{0, \max}^A = \rho(I - \Delta + B \mathrm{diag}(\ov x)\big(I - \Psi(\ov x)\big)W)\,,
\end{equation} 
where $\ov x$ in $[0,1]^{\V}$ are the upper bounds of $x(t)$ for all $t\geq0$.

\begin{theorem}\label{theo:free-equilibrium}
	Consider the adoption-opinion model \eqref{eq:vector_model} under Assumption \ref{ass:ass1}. Then, if $R_{0, \max}^A <1$, then the adoption-free equilibrium is globally asymptotically stable.
\end{theorem}\medskip
We can now consider the scenario when all communities are highly skeptical about the innovation and we can define
\begin{equation}\label{eq:r0_max}
	R_{0, \min}^A = \rho(I - \Delta + B \mathrm{diag}(\underline{x})\big(I - \Psi(\underline{x})\big)W)\,,
\end{equation}
where $\underline{x}$ in $[0,1]^{\V}$ is the vector of lower bounds of $x(t)$ for all $t\geq0$.
\MA{The following result gives sufficient conditions for the instability of the adoption-free equilibrium. In this case, even a small fraction of adopters can trigger spreading, driving the system toward regions of the phase space that are more relevant for our analysis.}
\begin{proposition}
	Consider the adoption-opinion model \eqref{eq:vector_model}. Then, if $R_{0, \min}^A >1$, then the adoption-free equilibrium is unstable.
\end{proposition}\medskip

\subsection{Adoption-diffused equilibrium}
Since we analyzed the stability of the adoption-free equilibrium, the next step is to study the adoption-diffused equilibria with $a^* > \0$. 
The following result characterizes conditions for its existence and stability. 
First define 
\be\label{eq:nu} \nu := \max_{j}\{ \theta_j(1-\underline{x}_j) - \delta_j \},	\ee
and for a given equilibrium $(a^*, d^*, x^*)$, let  
\begin{equation}\label{eq:varphi}
	\mspace{-5mu}\varphi := \mspace{-5mu} \max_{a,d,x \in [0,1]^{\V}} \!\! \left\| I \!-\! \Delta \!-\! \mc B^* \!\!\!+\! B\mathrm{diag}(x)\mathrm{diag}(\1-a-d) W \right\|_{\infty}\,,
\end{equation}
where $\mc B^* = B \mathrm{diag}\big( \mathrm{diag}(x(t)) W a^* \big)$.
\begin{theorem}\label{theo:diffused-equilibrium}
	Consider the adoption-opinion model \eqref{eq:vector_model}. If $R_{0, \min}^A >1$, then 
	\begin{enumerate}
		\item[(i)] there exists at least one adoption-diffused equilibrium $(a^*, d^*, x^*)$ with $a^*>\0$,
		\item[(ii)] if it also holds that
		\be \label{eq:hyp1}\beta_j \sum_{k \in \mc N_j} W_{jk} \leq \delta_j + \mc B^*_{jj}\,,\ee
		and there exist $\varsigma_1, \varsigma_2 >0$ such that 
		\begin{align}
			\nu^2 + \frac{\varsigma_2 \nu^2}{1 - \eta^2} + \frac{\varsigma_1 \varphi^2}{1 - \varphi^2} &< \varsigma_1 \,,\label{eq:hyp3} \\
			\rho^2(B^*) + \frac{\varsigma_1 \rho^2(B^*)}{1 - \varphi^2} + \frac{\varsigma_2 \eta^2}{1 - \eta^2}&< \varsigma_2\,, \label{eq:hyp4}
		\end{align}
		then the equilibrium $(a^*, d^*, x^*)$ is asymptotically stable for any non-zero initial condition.
	\end{enumerate}
\end{theorem}\medskip

It is worth noting that the coupled dynamics \eqref{eq:vector_model} exhibit significant complexity, and our results may not capture the entire picture. Theorem \ref{theo:diffused-equilibrium} provides only sufficient conditions for the convergence of the adoption-diffused equilibrium.\medskip

\section{Model Predictive control}\label{sec:mpc}
We aim now to introduce a control function $u(t)$ that facilitates the spread of innovation within the population.
\MA{One possible intervention strategy is to act on the opinion dynamics, aiming to steer individuals’ attitudes toward greater favorability for adoption. This can be implemented, for example, through awareness campaigns, targeted advertising, or information provision, which influence social interactions and, in particular, the perceived benefits of adopting the innovation. The controlled opinion dynamics can then be written as
	\vspace{-0.2cm}
	\be \label{eq:opinion-control}x(t+1)= (I-\Lambda-\Xi) (x(0) + u(t)) + \Lambda \tilde{W} x(t) + \Xi W a(t)\,.\ee
Alternative strategies could target adoption incentives or directly improve service quality; while potentially effective, these approaches are generally more costly or harder to implement. To account for real-world constraints, we impose a budget limit on the intervention effort: $1^T u(t) \leq C$ for all $t \ge 0$, where $C>0$ represents the available resources at each time step, ensuring that the control remains feasible.
}
To design effective interventions, we first consider an optimization problem where the control policy is determined by taking into account the system's asymptotic behavior. Specifically, we seek a constant optimal control policy $\bar{u}$ 
\MA{that minimizes a cost function balancing three competing objectives: maximizing the fraction of adopters, minimizing the fraction of dissatisfied individuals, and limiting the intervention effort itself. Formally, we define}
\begin{equation}\label{eq:opt-constant}
	\begin{aligned}
		\bar{u} \mspace{10mu} = \mspace{10mu} &\argmin_{u}{\quad  \sum_{j=1}^n \left[- Q^A_j (a_{c, j}^*)^2 + Q^D_{j} (d_{c, j}^*)^2+ L_j u_j^2\right]} \\[1ex]
		&\text{subject to: } 1^Tu \leq C\,, \quad u \in [0, 1-x(0)]\\[1ex]
		& \mspace{75mu} R_{0, \min}^A>1 \text{ and } \eqref{eq:hyp1}, \eqref{eq:hyp3}, \eqref{eq:hyp4},
	\end{aligned}
\end{equation}
where $a_{c, j}^*$ and $d_{c, j}^*$ denote the equilibrium values of adopters and dissatisfied individuals for each community $j$ for the model \eqref{eq:vector_model}, with the controlled opinion dynamics in \eqref{eq:opinion-control}. The weights $Q^A_j, Q^D_j, L_j$ encode the relative importance of maximizing adoption, minimizing dissatisfaction, and limiting intervention effort, respectively. 
\MA{Including a term like $L_j u_j^2$ is natural, as real-world interventions, such as campaigns or information provision, consume resources and effort; penalizing large interventions ensures that the optimal policy is both effective and feasible in practice. Finally, $\bar{u}$ corresponds to the optimal constant policy that guarantees stability of the adoption-diffused equilibrium.}

Beyond the steady-state optimization, we also consider a finite-horizon optimal control problem over a planning horizon $N > 0$. Given an initial state at time $t$, we seek a control policy $u(\cdot|t)$ that minimizes the total cost over the horizon while ensuring system feasibility. In this context, $a(k|t)$ represents the predicted evolution of adopters at time $k$ within the prediction horizon, given the system state at time $t$. Similarly, $d(k|t)$ and $x(k|t)$ denote the predicted dynamics of dissatisfied individuals and opinions, respectively.
Denote $F_1(\cdot)$, $F_2(\cdot)$ and $F_3(\cdot)$ as the right-hand side of the first equation in \eqref{eq:vector_model}, the second equation in \eqref{eq:vector_model} and \eqref{eq:opinion-control}, respectively. 
The optimal control problem is formulated as
\begin{equation} \label{eq:opt}
	\begin{aligned}
		\min_{U(t)} & \sum_{k=0}^{N-1} \sum_{j=1}^n \left[ - Q^A_j a_j^2(k|t) + Q^D_j d_j^2(k|t) + L_j u_j^2(k|t) \right] \\[1ex]
		\text{s.t. }&  1^T u(k|t) \leq C, \quad \forall k \in [0, N-1]\\
		& u(k|t) \in [0, 1 - x(0)], \quad \forall k \in [0, N-1] \\[1ex]
		& (a(0|t), \, d(0|t), \, x(0|t)) = (a(t),\, d(t),\, x(t)) \\
		& \begin{cases}
			a(k+1|t) = F_1(a(k|t), d(k|t), x(k|t)) \\
			d(k+1|t) = F_2(a(k|t), d(k|t), x(k|t)) \\
			x(k+1|t) = F_3(a(k|t), d(k|t), x(k|t)) 
		\end{cases} \\[2pt]
		& (a(N|t),\, d(N|t), \, x(N|t)) = (a_c^*,\, d_c^*,\, x_c^*),
	\end{aligned}
\end{equation}
where $a_{c}^*$, $d_{c}^*$ and $x_c^*$ denote the equilibrium values of adopters, dissatisfied individuals and opinions in the system \eqref{eq:vector_model}, incorporating the controlled opinion dynamics in \eqref{eq:opinion-control} with constant control $\bar{u}$ in \eqref{eq:opt-constant}. The set of controls over the prediction horizon is given by $U(t) = \{ u(0|t), \dots, u(N-1|t)\}$. 
The proposed MPC algorithm is outlined in Algorithm \ref{alg:1}.

\begin{algorithm}
	\caption{MPC algorithm} \label{alg:1}
	\begin{algorithmic}[1]
		\For{$t \geq 0$}
		\State Observe the current state $(a(t), d(t), x(t)) \in \mathcal{S}$.
		\State Solve problem \eqref{eq:opt} to obtain optimal control set $U^*(t)$.
		\State Apply the first control input: $u(t) = u^*(0|t)$ to the system \eqref{eq:vector_model}, with \eqref{eq:opinion-control}.
		\EndFor
	\end{algorithmic}
\end{algorithm}

In principle, solving \eqref{eq:opt} yields an optimal control sequence $U^*(t)$ over the prediction horizon of length $N$, which could theoretically be applied in its entirety. However, implementing the control policy in a receding horizon fashion, i.e., by setting $u(t) = u^*(0|t)$ and discarding the remaining elements of $U^*(t)$, is preferable. This approach accounts for external and uncontrollable factors that may temporarily affect the adoption dynamics, as well as the simplifying assumptions made in our model to incorporate such influences. 

\MA{We remark that MPC frameworks are typically accompanied by formal guarantees, namely recursive feasibility of the optimal control problem and asymptotic convergence of the trajectory to the desired equilibrium. Such properties were at present only verified numerically while Proving such properties for the controlled adoption–opinion model requires dedicated technical analysis, which is left for future work.}

\subsection{Simulation Results}
\begin{figure}
	\hspace{-4pt}
	\includegraphics[width=4.45cm, height=3.5cm]{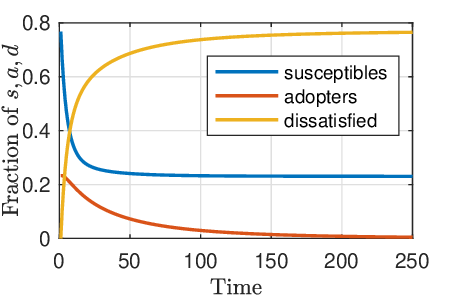}
	\hspace{-13pt}
	\includegraphics[width=4.45cm, height=3.5cm]{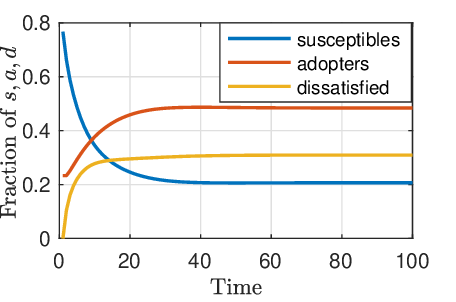}
	\caption{Numerical simulation of the aggregate uncontrolled dynamics \eqref{eq:vector_model} (left) and the corresponding solution under the MPC algorithm \ref{alg:1} (right).\\}
	\label{fig:mpc}
\end{figure}

\begin{figure}
	\hspace{-10pt}
	\includegraphics[width=4.5cm, height=6cm]{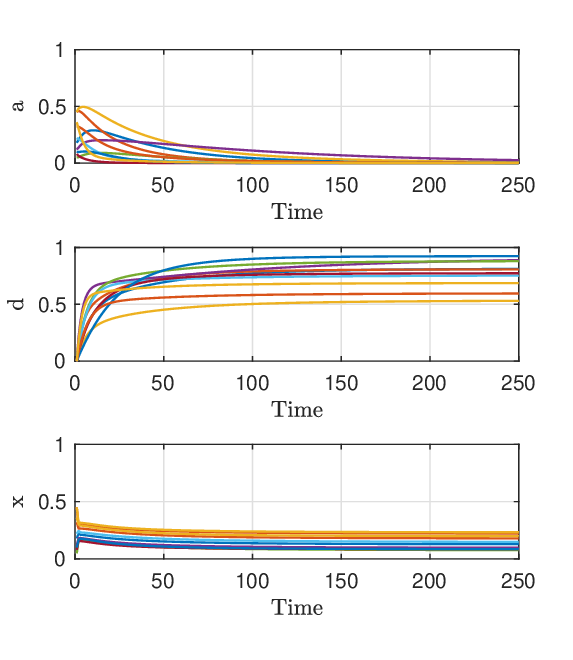}
	\hspace{-14pt}
	\includegraphics[width=4.5cm, height=6cm]{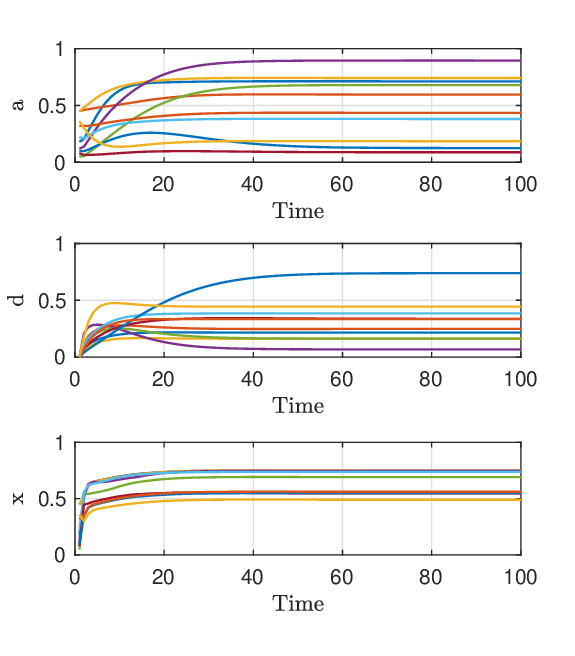}
	\caption{Numerical simulation of uncontrolled dynamics \eqref{eq:vector_model} (left) and the corresponding solution under the MPC algorithm \ref{alg:1} (right) in each community.}
	\label{fig:mpc2}
\end{figure}
In this section, we present numerical simulations to illustrate the coupled dynamics of innovation adoption and opinion formation among $n$ interacting communities over two distinct networks. Model parameters are randomly generated. The initial conditions represent an early stage of innovation diffusion, where only a few individuals have adopted the innovation, and there are no dissatisfied individuals.
The algorithms are performed using a sequential quadratic programming (SQP) method, in which a quadratic programming (QP) subproblem is solved at each iteration. 
Figure \ref{fig:mpc} shows the evolution of the aggregate adoption dynamics, summed over $n=10$ communities: on the left the uncontrolled fractions of susceptibles, adopters and dissatisfied individuals across the whole network are reported, while the right-hand plot represents the solution of the Model Predictive Control (MPC) algorithm \ref{alg:1}. The uncontrolled fraction of adopters decays to $0$ over time; however, when the MPC algorithm is applied, the solution maintains a positive fraction of adopters. This indicates that, although strategically influencing opinions may incur costs, the nudge control can be an effective way to sustain the adoption of innovation. Figure \ref{fig:mpc2} illustrates the evolution of adoption and opinion dynamics in each community. On the left, it presents the uncontrolled fractions of adopters, dissatisfied individuals, and opinions, while on the right, it shows the MPC solution for each community.

\begin{figure}
	\centering
	\includegraphics[scale=0.64]{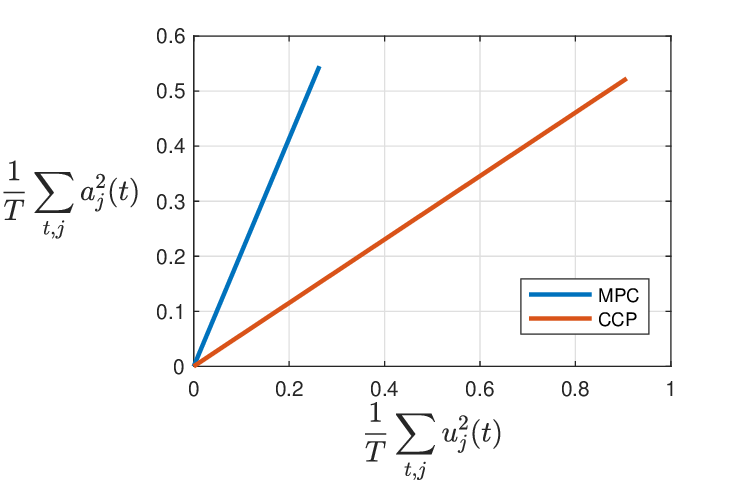}
	\caption{Control cost vs effectiveness for constant control policy (CCP) \eqref{eq:opt-constant} (in red) and MPC algorithm \ref{alg:1} (in blue).}
	\label{fig:pareto}
\end{figure}

To assess the trade-off between effectiveness and implementation effort, we introduce two performance indicators comparing the constant control policy (CCP) \eqref{eq:opt-constant} with the one derived from the MPC algorithm \ref{alg:1}. The first metric quantifies the total fraction of adopters, that is the effectiveness of a policy in promoting innovation adoption over the considered time horizon. A higher value of this indicator implies a more successful diffusion strategy.  
However, evaluating policy impact requires also considering the effort needed for its implementation. To this end, we define a second indicator, the control cost, which measures the intervention effort required from policymakers or stakeholders.  
Figure \ref{fig:pareto} compares the two policies based on these indicators, highlighting the balance between diffusion effectiveness and control effort. The results highlight that the MPC algorithm outperforms the constant policy, achieving a positive fraction of adopters at a lower cost. This advantage stems from the ability of MPC to predict and optimize adoption dynamics over a given planning horizon, unlike the constant policy, which lacks such foresight.

\section{Conclusion}\label{sec:conclusion} 
We presented an adoption-opinion model that couples opinion dynamics with a compartmental adoption framework on a multilayer network to study the diffusion of sustainable behaviors. We analyzed equilibrium points stability and formulated an optimal control problem, introducing a MPC framework that optimizes interventions by shaping opinions rather than directly enforcing adoption. This nudge-based control strategy allows policymakers to influence diffusion indirectly, making interventions more effective and scalable. Numerical simulations demonstrate that, in the absence of control, adoption stagnates, whereas MPC-driven interventions sustain and enhance adoption across communities and it outperforms the constant policy. 
\MA{A key direction for future work is to systematically compare these different approaches, in order to understand where interventions are most effective and cost-efficient. Another avenue is to design policies that ensure permanent changes in adoption dynamics, extending the impact of interventions beyond the controlled horizon.}

%
\bibliographystyle{IEEEtran}
\bibliography{bib}

\end{document}